\documentclass[12pt]{iopart}

\usepackage{graphicx}

\begin{document}
\title[ ]{Spin-polarized current separator based on a fork-shaped Rashba nanostructure}

\author{Xianbo Xiao$^{1,2}$ and Yuguang Chen$^{3\dag}$
\footnotetext{Author to whom any correspondence should be addressed. } }

\address{$^1$ School of Computer, Jiangxi University of Traditional Chinese Medicine, Nanchang 330004, China.\\
$^2$ Institute for Advanced Study, Nanchang University, Nanchang 330031, China.\\
$^3$ Department of Physics, Tongji University, Shanghai 200092, China.}
\ead{ygchen@tongji.edu.cn}

\begin{abstract}

A scheme for a spin-polarized current separator is proposed by
studying the spin-dependent electron transport of a fork-shaped
nanostructure with Rashba spin-orbit coupling (SOC), connected to
three leads with the same width. It is found that two spin-polarized
currents are of the same magnitude but opposite polarizations can be
generated simultaneously in the two output leads when the
spin-unpolarized electrons injected from the input lead. The
underlying physics is revealed to originate from the different
spin-dependent conductance caused by the effects of Rashba SOC and
the geometrical structure of the system. Further study shows that
the spin-polarized current with strong a robustness against
disorder, demonstrates the feasibility of the proposed nanostructure
for a real application.

\end{abstract}

\maketitle

\newpage

\section{INTRODUCTION}

In the past decades, spin-dependent electron transport in
semiconductor nanostructures has drawn unprecedented attention
because of its potential applications to semiconductor
spintronics,$^1$ in which the electron spin rather than its charge
is utilized for information processing. One of the primary tasks in
the development of semiconductor spintronics is to be capable of
generating and manipulating excess spin in semiconductor
nanostructures, particularly by all electrical means. The Rashba
spin-orbit coupling (SOC),$^2$ existing in asymmetric
heterostructures and can be controlled by an external gate
voltage,$^{3,4}$ may be an efficient method to satisfy this goal.

Various spin filtering devices have been proposed based on the
Rashba semiconductor nanostructures without need for a magnetic
element or an external magnetic field, such as T-shape electron
waveguide,$^{5-9}$ quantum wires,$^{10-14}$ wire network,$^{15}$
two-dimensional electron gas (2DEG),$^{16}$ and quantum
rings.$^{17-18}$ Recently, an interesting Fano-Rashba effect has
been found in a straight quantum wire with local Rashba SOC.$^{19}$
This effect is attributed to the interference between the bound
states formed by the Rashba SOC and the electrons in the conduction
channel, giving rise to pronounced dips in the linear charge
conductance. Apart from the SOC-induced bound states, the effects of
structure-induced bound states on the electron and spin transport
have also been concerned intensively.$^{6,20,21}$ In our recent
works, we have investigated the spin-polarized electron transport
properties of several typical Rashba quantum wires and found that
they are very sensitive to the systems' longitudinal symmetry.
Spin-polarized current can be generated in the longitudinally
asymmetry systems when spin-unpolarized injections. Especially, the
magnitudes of the spin polarization around the structure-induced
Fano resonances are very large.$^{22,23}$ However, in the
longitudinally symmetrical system no spin-polarized current can be
achieved, despite the existence of the SOC- or/and structure-induced
bound states.$^{24}$

Inspired by the three works above, in this paper, we study the
spin-dependent electron transport for a fork-shaped Rashba
nanostructure with longitudinal-inversion symmetry. It is shown that
two spin-polarized currents, with the same magnitude but different
polarized directions, can be achieved in the two output leads in
spite of spin-unpolarized injections and they still survive even in
the presence of strong disorder. Therefore, a spin-polarized current
separator device can be devised by using this system. The rest of
this paper is organized as follows. In Section II, the theoretical
model and analysis are presented. In Section III, the numerical
results are illustrated and discussed. A conclusion is given in
Section IV.

\section{MODEL AND ANALYSIS}

The investigated system in present work is schematically depicted in
Fig. 1, where a 2DEG in the $(x,y)$ plane is restricted to a
fork-shaped nanostructure by a confining potential $V(x,y)$. The
2DEG is confined in an asymmetric quantum well, where the Rashba SOC
is assumed to play a dominantly role. The nanostructure consists of
three narrow regions and a wide region. The wide region has a length
$L_{2}$ and a uniform width $W_{2}$, while the narrow region has a
length $L_{1}$ and a uniform width $W_{1}$, connected to a
semi-infinite lead with the same width. The three connecting leads
are normal-conductor electrodes without SOC, since we are only
interested in spin-unpolarized injection. Such kind of Rashba system
can be described by the spin-resolved discrete lattice model. The
tight-binding Hamiltonian including the Rashba SOC on a square
lattice is given as follow,
\begin{eqnarray}
H=H_0+H_{so}+V,
\end{eqnarray}
where
\begin{eqnarray}
H_0=\sum\limits_{lm\sigma}\varepsilon_{lm\sigma}c_{lm\sigma}^{\dag}c_{lm\sigma}-t\sum\limits_{lm\sigma}\{c_{l+1m\sigma}^{\dag}c_{lm\sigma}\nonumber\\
+c_{lm+1\sigma}^{\dag}c_{lm\sigma}+H.c\},
\end{eqnarray}
\begin{eqnarray}
H_{so}=t_{so}\sum\limits_{lm\sigma\sigma'}\{c_{l+1m\sigma'}^{\dag}(i\sigma_{y})_{\sigma\sigma'}c_{lm\sigma}\nonumber\\
-c_{lm+1\sigma'}^{\dag}(i\sigma_{x})_{\sigma\sigma'}c_{lm\sigma}+H.c\},
\end{eqnarray}
and
\begin{eqnarray}
V=\sum\limits_{lm\sigma}V_{lm}c_{lm\sigma}^{\dag}c_{lm\sigma},
\end{eqnarray}
in which $c_{lm\sigma}^{\dag}(c_{lm\sigma})$ is the creation
(annihilation) operator of electron at site $(lm)$ with spin
$\sigma$, $\sigma_{x(y)}$ is Pauli matrix, and
$\varepsilon_{lm\sigma}=4t$ is the on-site energy with the hopping
energy $t=\hbar^{2}/2m^{\ast}a^{2}$, here $m^{\ast}$ and $a$ are the
effective mass of electron and lattice constant, respectively.
$V_{lm}$ is the additional confining potential. The SOC strength is
$t_{so}=\alpha/2a$ with the Rashba constant $\alpha$. The Anderson
disorder can be introduced by the fluctuation of the on-site
energies, which distributes randomly within the range width $w$
[$\varepsilon_{lm\sigma}= \varepsilon_{lm\sigma}+w_{lm}$ with
$-w/2<w_{lm}<w/2$].

In the ballistic transport, the spin-dependent conductance is
obtained from the Landauer-B$\ddot{u}$ttiker formula$^{25}$ with the
help of the nonequilibrium Green function formalism.$^{26}$ In order
to calculate the Green function of the whole system conveniently,
the tight-binding Hamiltonian (1) is divided into two parts in the
column cell
\begin{eqnarray}
H=\sum\limits_{l\sigma\sigma'}H_{l}^{\sigma\sigma'}+\sum\limits_{l\sigma\sigma'}(H_{l,l+1}^{\sigma\sigma'}+H_{l+1,l}^{\sigma'\sigma}),
\end{eqnarray}
where $H_{l}^{\sigma\sigma'}$ is the Hamiltonian of the $l$th
isolated column cell, $H_{l,l+1}^{\sigma\sigma'}$ and
$H_{l+1,l}^{\sigma'\sigma}$ are intercell Hamiltonian between the
$l$th column cell and the $(l+1)$th column cell with
$H_{l,l+1}^{\sigma\sigma'}=(H_{l+1,l}^{\sigma'\sigma})^\dag$. The
Green function of the whole system can be computed by a set of
recursive formulas,
\begin{eqnarray}
\langle l+1|G_{l+1}|l+1\rangle^{-1}=E-H_{l+1}-H_{l+1,l}\langle l|G_{l}|l\rangle H_{l,l+1},\nonumber\\
\langle l+1|G_{l+1}|0\rangle=\langle l+1|G_{l+1}|l+1\rangle
H_{l+1,l}\langle l| G_{l}|0\rangle,
\end{eqnarray}
where $\langle l+1|G_{l+1}|l+1\rangle$ and $\langle
l+1|G_{l+1}|0\rangle$ are respectively the diagonal and off-diagonal
Green function, and
\begin{eqnarray}
H_{l+1}=\left(
\begin{array}{cc}
H_{l+1}^{\sigma\sigma} & H_{l+1}^{\sigma\sigma'} \\
H_{l+1}^{\sigma'\sigma} & H_{l+1}^{\sigma'\sigma'} \end{array}
\right),~~ H_{l+1,l}=(H_{l,l+1})^\dag=\left(
\begin{array}{cc}
H_{l+1,l}^{\sigma\sigma} & H_{l+1,l}^{\sigma\sigma'} \\
H_{l+1,l}^{\sigma'\sigma} & H_{l+1,l}^{\sigma'\sigma'} \end{array}
\right).
\end{eqnarray}

Utilizing the Green function of the whole system obtained above, the
spin-dependent conductance from arbitrary lead $p$ to lead $q$ is
given by
\begin{eqnarray}
G^{\sigma\sigma'}_{pq}=e^2/hTr[\Gamma_{p}^{\sigma}G^{r}\Gamma_{q}^{\sigma'}G^{a}],
\end{eqnarray}
where $\Gamma_{p(q)}=i[\sum_{p(q)}^{r}-\sum_{p(q)}^{a}]$ with the
self-energy from the lead
$\sum_{p(q)}^{r}=(\sum_{p(q)}^{a})^{\ast}$, the trace is over the
spatial and spin degrees of freedom. $G^{r}(G^{a})$ is the retarded
(advanced) Green function of the whole system, which can be computed
by the spin-resolved recursive Green function method,$^{23}$  and
$G^{a}=(G^{r})^\dag$.

In the following calculation, the structural parameters of the
system are fixed at $L_1=L_2=10~a$, $W_1=10~a$, and $W_2=40~a$. All
the energy is normalized by the hoping energy $t(t=1)$. And the $z$
axis is chosen as the spin-quantized axis so that
$|\uparrow>=(1,0)^{T}$ represents the spin-up state and
$|\downarrow>=(0,1)^{T}$ denotes the spin-down state, where $T$
means transposition. For simplicity, the hard-wall confining
potential approximation is adopted to determine the boundary of the
nanostructure since different confining potentials only alter the
positions of the subbands and the energy gaps between them. The
charge conductance and the spin conductance of $z$-component are
defined as
$G^e_{pq}=G^{\uparrow\uparrow}_{pq}+G^{\uparrow\downarrow}_{pq}+G^{\downarrow\downarrow}_{pq}+G^{\downarrow\uparrow}_{pq}$
and
$G^{Sz}_{pq}=\frac{e}{4\pi}\frac{G^{\uparrow\uparrow}_{pq}+G^{\downarrow\uparrow}_{pq}-G^{\downarrow\downarrow}_{pq}-G^{\uparrow\downarrow}_{pq}}{e^2/h}$,
respectively. Here the charge conductance means the transfer
probability of electrons, and the spin conductance represents the
change in local spin density between the input lead and the output
lead caused by the transport of spin-polarized electrons.$^{27}$

\section{RESULTS AND DISCUSSION}

In our numerical example, we choose the same material as that in
Ref. [23], where the requirements of the parameters have been
discussed. Figure 2 shows the electron energy ($E$) dependence of
the charge and spin conductance when the spin-unpolarized electron
injected from lead 1. The Rashba SOC strength $t_{so}=0.19$. The
step-like structures, oscillation caused by interference, and
SOC-induced Fano resonance dips (see the red circles in Fig. 2(a))
can be found in the charge conductance. In addition, due to the
system is longitudinally symmetrical, electrons have the same chance
be transmitted to the different output leads. Therefore, as shown in
Fig. 2(a) and (b), the charge conductance from lead 1 to 2 is the
same as that from lead 1 to 3.  However, the corresponding spin
conductance from lead 1 to 2 is quite different from that from lead
1 to 3, as depicted in Fig. 2(c), the magnitudes of the spin
conductance from the injecting lead to the two outgoing leads are
always equal but their signs are contrary. In particular, a very
large spin-polarized current can be generated at the
structure-induced Fano resonances (such as $E=0.16,~0.44,$ etc.). It
has demonstrated in our previous papers$^{22,23}$ that the magnitude
of this spin-polarized current can be tuned by both the strength of
Rashba SOC and the structural parameters of the system so that we do
not presented these results here.

The remarkable difference in the spin conductance between the upper
and lower output leads can be utilized to design a spin-polarized
current separator, i.e., if a spin-polarized current generated in
one output lead, there must be another one in possession of the same
magnitude but adverse polarized directions achieved in the other
output lead. The physical mechanism of this device owing to the
effect of the Rashba SOC and the geometrical structure of the
system. The spin-dependent conductance from the input lead 1 to the
output leads 2 and 3 as function of the electron energy is
illustrated in Fig. 3(a) and (b), respectively. The strength of
Rashba SOC is the same as that in Fig. 2. The fork-shaped Rashba
nanostructure can be equivalently viewed as two coupled zigzag wire,
whose longitudinal and transversal symmetries are broken.$^{12}$ So
the relations
$G^{\uparrow\uparrow}_{12(3)}=G^{\downarrow\downarrow}_{12(3)}$ and
$G^{\uparrow\downarrow}_{12(3)}=G^{\downarrow\uparrow}_{12(3)}$
cannot be guaranteed, as shown in Fig. 3, leading to the nonzero
spin conductance (see Fig. 2(c)) in respective lead. Furthermore,
because the two output leads 2 and 3 lie symmetrically in the
opposite direction with respect to the input lead 1, the total
current must still be spin-unpolarized.$^{12,24}$ Therefore, the
transmission probability of the spin-up (-down) electron from lead 1
to 2 always equals that of the spin-down (-up) electron from lead 1
to 3, that is, $G^{\sigma\sigma}_{12}=G^{\sigma'\sigma'}_{13}$ and
$G^{\sigma\sigma'}_{12}=G^{\sigma'\sigma}_{13}$. As a consequence,
the signs of the spin conductance from the lead 1 to leads 2 and 3
are contrary all along.

The above proposed spin-polarized current separator is based upon a
perfectly clean system, where no elastic or inelastic scattering
happens. However, in a realistic system, there are many impurities
in the sample. The impurities in any semiconductor heterostructure
may induce a random Rashba field, which gives rise to many new
effects such as the realization of the minimal possible strength of
SOC$^{28}$ and the localization of the edge electrons for
sufficiently strong electron-electron interactions.$^{29}$ Thus the
effect of disorder should be considered in practical application.
The spin conductance from the input lead to the different output
leads as function of the electron energy for (weak and strong)
different disorders $w$ are plotted in Fig. 4. The SOC strength is
also set as $t_{so}=0.19$. The spin conductance is destroyed when
the impurities exist in the system and its magnitude become smaller
with the increase of disorder. However, as shown in Fig. 4(c), the
magnitude of the spin conductance around the structure-induced Fano
resonances is still large when the disorder strength $w=0.6$, which
means that a comparatively large spin-polarized current can be
obtained in the output leads even in the presence of strong
disorder.

\section{CONCLUSION}

In conclusion, a scheme of a spin-polarized current separator is
proposed by investigating the spin-dependent electron transport of a
fork-shaped nanostructure under the modulation of the Rashba SOC.
Two spin-polarized currents with the same magnitude but different
polarizations can be generated synchronously in the two output leads
due to the distinct spin-dependent conductance results from the
effects of SOC and the geometrical structure. The opposite
spin-polarized currents can be generated and controlled by
electrical means and they are robust against disorder. Thus the
proposed nanostructure does not require the application of magnetic
fields, external radiation or ferromagnetic leads, and has great
potential for real applications.

\begin{flushleft}
\section*{ACKNOWLEDGMENT}
\end{flushleft}

This work was supported by the National Natural Science Foundation
of China under Grant No. 10774112.

\section*{References}

~~~~~~$^1$D. D. Awschalom, D. Loss, and Samarth N. \emph{Semiconductor
Spintronics and Quantum Computation}, (Springer, Berlin) 2002; I.
Zutic, J. Fabian, and S. D. Sarma, Rev. Mod. Phys. {\bf76} 323
(2004); J. Fabian, A. Matos-Abiague, C. Ertler, P. Stano, and I.
Zutic, Acta Phys. Slov. {\bf57} 565 (2007), and references
therein.\\

$^2$Y. A. Bychkov and E. I. Rashba, J. Phys. C \textbf{17} 6039
(1984).\\

$^3$D. Grundler, Phys. Rev. Lett. \textbf{84} 6074 (2000).\\

$^4$T. Koga, J. Nitta, T. Akazaki, and H. Takayanagi, Phys. Rev.
Lett. \textbf{89} 046801 (2002).\\

$^5$M. Yamamoto, T. Ohtsuki, and B. Kramer,Phys. Rev. B \textbf{72}
115321 (2005).\\

$^{6}$F. Zhai and H. Q. Xu, Phys. Rev. B \textbf{76} 035306
(2007).\\

$^{7}$S. Bellucci and P. Onorato, Phys. Rev. B \textbf{77} 075303
(2008).\\

$^{8}$M. Yamamoto and B. Kramer, J. Appl. Phys., \textbf{103}
123703 (2008).\\

$^{9}$T. Yokoyama and M. Eto, Phys. Rev. B \textbf{80} 125311
(2009).\\

$^{10}$J.Ohe, M. Yamamoto, T. Ohtsuki, and J. Nitta, Phys. Rev. B
\textbf{72} 041308(R) (2005).\\

$^{11}$Q. F. Sun and X. C. Xie, Phys. Rev. B \textbf{71} 155321
(2005).\\

$^{12}$Z. Y. Zhang, J. Phys: Condens. Matter \textbf{19} 016209
(2007).\\

$^{13}$G. H. Liu and G. H. Zhou, J. Appl. Phys. \textbf{101} 063704
(2007).\\

$^{14}$G. I. Japaridze, H. Johannesson, and A. Ferraz, Phys. Rev. B
\textbf{80} 041308(R) (2009).\\

$^{15}$H. X.Wang, S. J. Xiong, and S. N. Evangelou, Phys. Lett. A
\textbf{356} 376 (2006).\\

$^{16}$A. P$\acute{a}$lyi, C. P$\acute{e}$terfalvi, and J. Cserti,
Phys. Rev. B \textbf{74} 073305 (2006).\\

$^{17}$P. F$\ddot{o}$ldi, O. K$\acute{a}$lm$\acute{a}$n, M. G.
Benedict, and F. M. Peeters, Phys. Rev. B \textbf{73} 155325 (2006).

$^{18}$P. F$\ddot{o}$ldi, O. K$\acute{a}$lm$\acute{a}$n, M. G.
Benedict, and F. M. Peeters, Nano Lett. \textbf{8} 2556 (2008).\\

$^{19}$D. S$\acute{a}$nchez and L. Serra, Phys. Rev. B \textbf{74}
153313 (2006).\\

$^{20}$Y. P. Chen, X. H. Yan, and Y. E Xie, Phys. Rev. B \textbf{71}
245335 (2005).\\

$^{21}$Y. P. Chen, Y. E Xie, and X. H. Yan, Phys. Rev. B \textbf{74}
035310 (2006).\\

$^{22}$X. B. Xiao, X. M. Li, and Y. G. Chen, Phys. Lett. A
\textbf{373} 4489 (2009).\\

$^{23}$X. B. Xiao and Y. G. Chen, Europhys. Lett.
\textbf{90} 47004 (2010).\\

$^{24}$X. B. Xiao, X. M. Li, and Y. G. Chen, ACTA PHYSICA SINICA
\textbf{58} 7909 (2009).\\

$^{25}$M. B$\ddot{u}$ttiker, Phys. Rev. Lett. \textbf{57} 1761
(1986).\\

$^{26}$L. W. Molenkamp, G. Schmidt, and G. E.W. Bauer, Phys. Rev. B
\textbf{62} 4790 (2000); T. P. Pareek and P. Bruno, Phys. Rev. B
\textbf{63} 165424-1 (2001).\\

$^{27}$D. V. Khomitsky, Phys. Rev. B \textbf{79} 205401 (2009).\\

$^{28}$E. Ya. Sherman, Phys. Rev. B \textbf{67} 161303(R) (2003).\\

$^{29}$A. Str$\ddot{o}$m, H. Johannesson and G. I. Japaridze, Phys.
Rev. Lett. \textbf{104} 256804 (2010).

\newpage

\begin{figure}
\center
\includegraphics[width=4.5in]{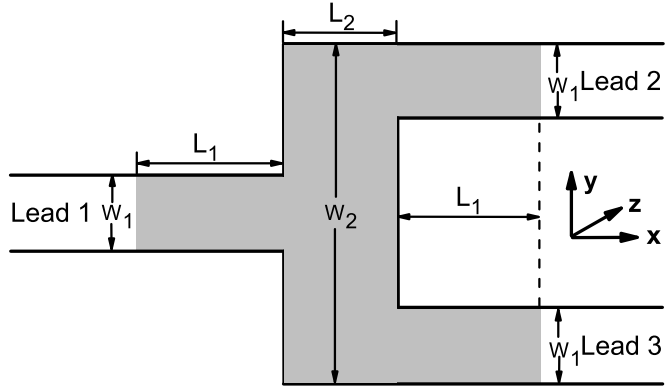}
\caption{Schematic diagram of the fork-shaped nanostructure with
Rashba SOC. The narrow regions have the same length $L_{1}$ and
width $W_{1}$, while the wide region has another length $L_{2}$ and
width $W_{2}$.}
\end{figure}

\begin{figure}
\center
\includegraphics[width=4.5in]{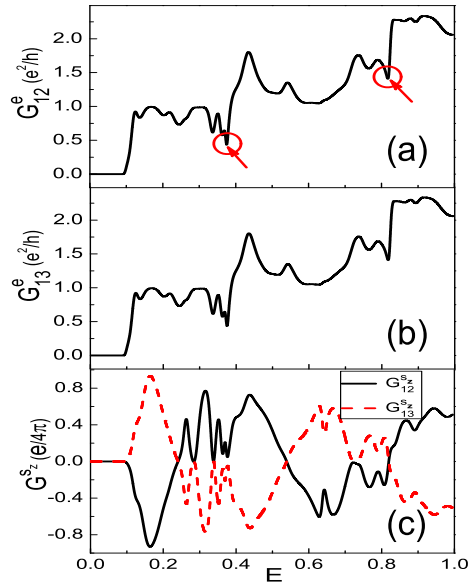}
\caption{(Color online) Conductance spectra of a fork-shaped Rashba
nanostructure as function of the electron energy for
spin-unpolarized electron injections: (a) the charge conductance
from lead 1 to 2; (b) the charge conductance from lead 1 to 3; (c)
the corresponding spin conductance from lead 1 to 2 (the solid line)
and 3 (the dash line). The Rashba SOC strength $t_{so}=0.19$.}
\end{figure}

\begin{figure}
\center
\includegraphics[width=4.0in]{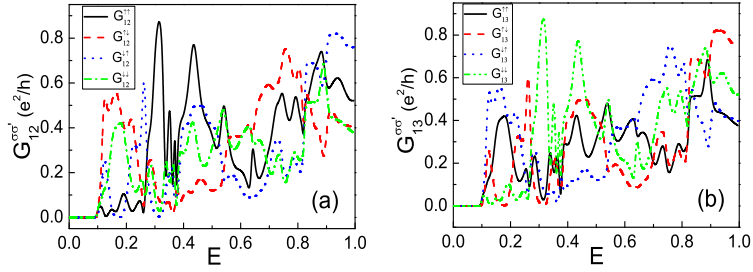}
\caption{(Color online) The calculated spin-dependent conductance as
function of the electron energy when the spin-unpolarized electron
travels from lead 1 to 2 (a) and 3 (b). The Rashba SOC strength is
the same as that in Fig. 2.}
\end{figure}

\begin{figure}
\center
\includegraphics[width=5.0in]{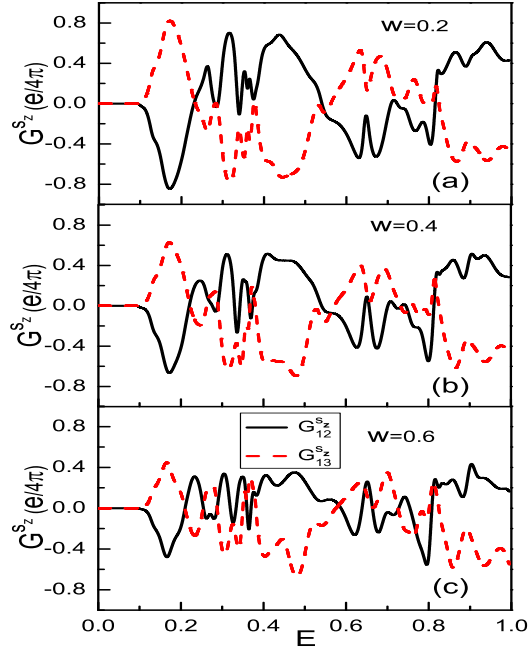}
\caption{(Color online) The calculated spin conductance as function
of the electron energy for different disorder strengths. The solid
line represents $G^{Sz}_{12}$ and the dashed line $G^{Sz}_{13}$. The
Rashba SOC strength is the same as that in Fig. 2.}
\end{figure}

\end{document}